\documentclass[twocolumn,showpacs,preprintnumbers,aps,prb,amsmath,amssymb]{revtex4}
\usepackage{graphicx}
\usepackage{dcolumn}
\usepackage{bm}
\begin{document}
\preprint{Prepared for PRB}
\begin{titlepage}
\title{Selective adsorption of first-row atoms on boron nitride nanotubes: the effect of localized states}
\author{Jia Li, Gang Zhou, Haitao Liu, and Wenhui Duan\footnote{Author to whom correspondence should be
addressed.\\ Email address: dwh@phys.tsinghua.edu.cn}}
\affiliation{Department of Physics, Tsinghua University, Beijing
100084, People's Republic of China}
\date{\today}

\begin{abstract}
First-principles calculations reveal that the adsorption of
representative first-row atoms with different electronegativity,
such as lithium (Li), carbon (C) and fluorine (F), on zigzag
single-walled boron nitride nanotubes (BNNTs) exhibits surprising
selectivity. The adsorption energy and adsorption site are
dependent upon the chemical activity of adsorbate with respect to
the B and N atoms in the host tube. In detail, the F atom prefers
to be adsorbed on the top of the B atom, the C atom is
energetically favorable to stay on the bridge site which is
perpendicular to the tube-axis, and the Li atom hardly adheres to
the tube (an endothermal reaction). The adsorption behavior of
these three types of elements on BNNTs is elucidated based on the
frontier molecular orbital theory. In addition, the mechanism of
modification of electronic structures of BNNTs by adsorption is
probed, and a feasible approach is proposed to tailor the
electronic properties of BNNTs.

\end{abstract}

\maketitle

\end{titlepage}

\section{Introduction}
Boron nitride nanotubes (BNNTs) have attracted an enormous amount
of
attention\cite{Chopra,Chen,Blase1,Rubio,Blase,Golberg,Hanjacs,Tangcc}
as a typical representative of III-V compound tubes, partially
because they have the morphology of honeycomb analogous to carbon
nanotubes (CNTs). However, different from CNTs, BNNTs are
semiconducting with an uniform wide energy gap,\cite{Blase1} and
their electronic properties are independent of the tube diameter,
chirality and of whether a nanotube is single-walled or
multiwalled.\cite{Rubio} Owing to the unique geometrical structure
and uniform semiconducting behavior, BNNTs are expected to have
significant applications in the molecule-based logic gates and
high-strength fibers.

On the other hand, for the hollow nanotube with high
surface-area-to-volume ratio, the doping is recognized to be a
significant and attractive approach\cite{Doping} of
functionalizing the nanotube because it provides various
possibilities for controlling the physical properties. In
particular, previous studies showed that the doping of carbon (C)
atoms into BNNTs could visibly decrease the band
gap,\cite{Golberg,Blase} and such a gap decrease is sensitively
dependent upon the content of C atoms.\cite{Blase} Subsequently,
Han {\it et al.} developed a simple chemical route to successfully
coat BNNTs with a conductive stannic oxide for the sensor
application.\cite{Hanjacs} Recently, Tang {\it et al.} found
experimentally that the resistance and resistivity of the
fluorinated BNNT are about 3 orders of magnitude less than those
of the pure BNNT, and suggested that it might be important for
applications in the future nanoscale electronic devices with
tunable properties.\cite{Tangcc} Thus, uniformly doped BNNTs
obtained through chemical modification could be very effective for
tailoring the electronic properties, and the related mechanism and
behavior are valuable to be explored in theory.

In this paper, we demonstrate a feasible way to modify the
electronic properties by atomic adsorbing on perfectly fabricated
BNNTs, which is different from the conventional way by doping
foreign species into host systems in the nanotube growth
process.\cite{Golberg,Blase,Tangcc} Firstly, we study the
adsorption of several representative first-row atoms with
different electronegativity on single-walled BNNTs through
first-principles calculations. And then we analyze the adsorption
selectivity of BNNTs from the localized characteristics of
electronic states near the Fermi level and discuss the adsorption
behavior of these atoms on BNNTs based on the frontier molecular
orbital theory. Finally we explore the mechanism of modification
of the electronic properties of BNNTs by adsorption.

\section{Model and method}
Since many experiments have shown that BNNTs prefer a nonhelical
or zigzag orientation during the growth,\cite{Zigzag} we
emphatically address the adsorption of zigzag BNNTs. A
finite-length cylindrical BN cage, consisting of 48 B and 48 N
atoms, is chosen to represent the stem of a zigzag single-walled
(8,0) BNNT. Two mouths of the tube are terminated by the hydrogen
atoms to avoid the boundary effect. In general, the interaction
between two different atoms or elements could be simply evaluated
in term of the Pauli electronegativity (PE) and first ionization
potential (FIP). With respect to PEs and FIPs of B and N atoms, we
choose three types of first-row atoms, lithium (Li), C and
fluorine (F), with different PE and FIP as the representatives of
adsorbate to study the adsorption properties of BNNTs.

\begin{figure} [tbp]
\includegraphics[width=0.5\textwidth]{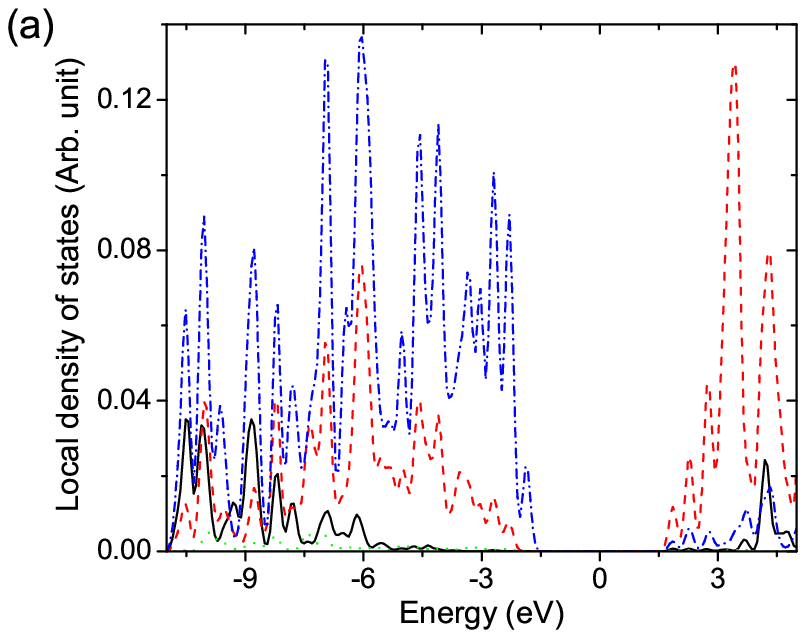}
\includegraphics[width=0.5\textwidth]{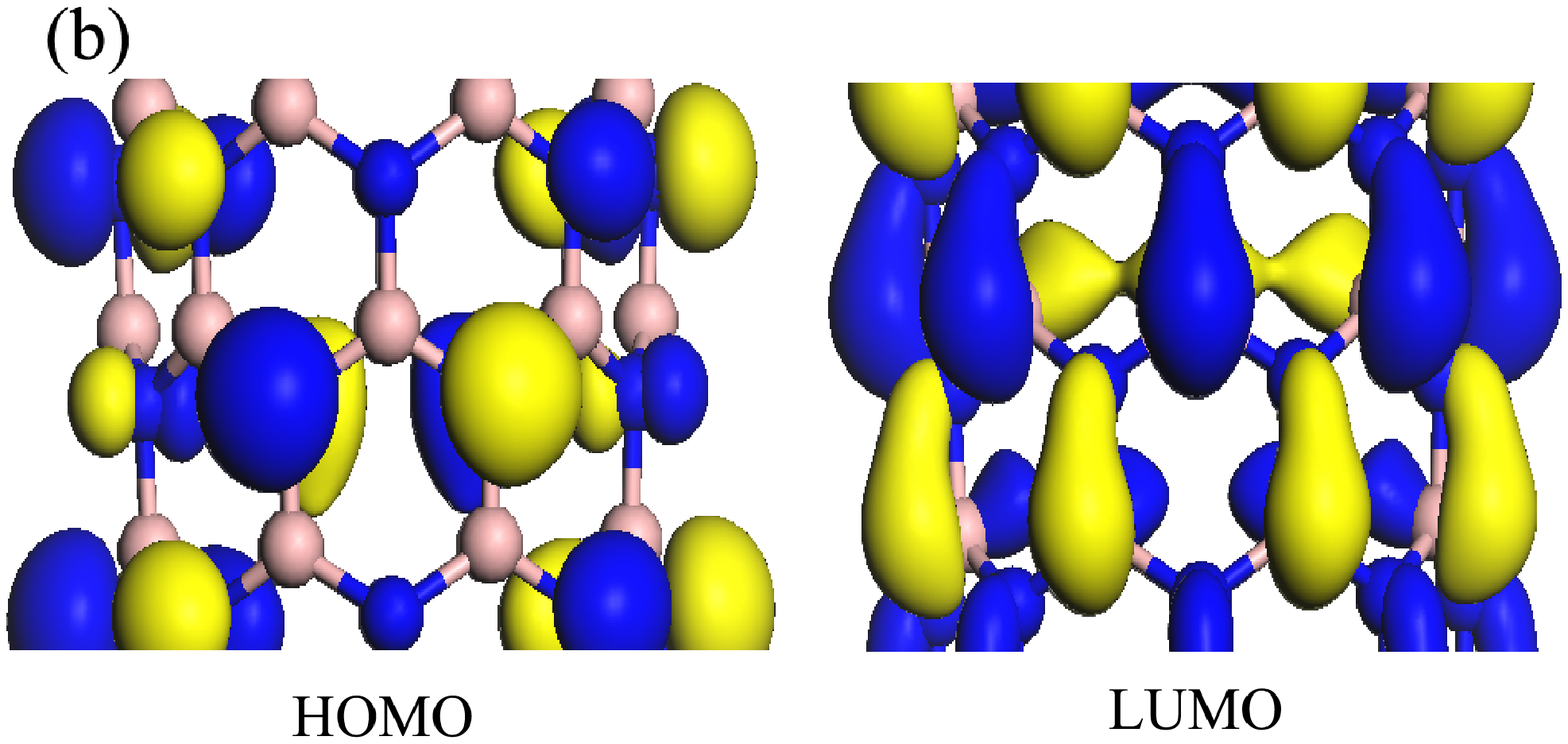}
\caption{\label{fig:fig1}(Color online) (a) Local density of
states of B and N atoms of the (8,0) BNNT. Black solid, red
dashed, green dotted and blue dash-dotted lines correspond to
B-$2s$, B-$2p$, N-$2s$ and N-$2p$, respectively. The Fermi level
is set to zero energy. (b) The HOMO (left) and LUMO (right) of the
(8,0) BNNT stem. Blue and pink balls correspond to N and B atoms,
respectively. The isosurfaces of the HOMO and LUMO at the values
of $0.02$ and $-0.02$ a.u. are depicted by the blue and yellow,
respectively.}
\end{figure}

The local-orbital density-functional method, DMol${^3}$, is
employed to carry out structural optimizations and total-energy
calculations.\cite{Delley} The generalized gradient approximation
combining the Perdew-Wang correlation functional \cite{Perdew}
with the Becke exchange functional \cite{Becke} is utilized. The
single-particle wave functions in the Kohn-Sham equations are
expanded by the double-numerical basis. Structural optimizations
are deemed sufficiently converged when the forces on all ions are
less than 10$^{-4}$ a.u..

\section{Results and discussion}

An important and primary feature of compound tubes is that the
distributions of positive and negative charges along the tube stem
are much more nonuniform than those of elementary tubes (e.g.,
CNTs) because of the unambiguous charge transfer from one
component with low electronegativity to another with high
electronegativity in tubes. Typically, the calculated charge
transfer from the B to N atom is 0.111 e per B-N pair in the (8,0)
BNNT. So in general the isolated BNNT could be regarded as an
unique tube that consists of the equal number of ``N cations'' and
``B anions'', and possesses the large ``ionicity''. In principle,
the resulting influences on the structures and properties of BNNTs
could be understood in virtue of the local density of states
(LDOS).

Fig.~\ref{fig:fig1}(a) shows the calculated LDOS of B and N atoms
of the (8,0) BNNT. It can be seen that $2s$ and $2p$ states from
the B and N atoms strongly mix together in the lower or higher
energy region far from the Fermi level. While the occupied and
unoccupied states near the Fermi level are mainly contributed by
$2p$ states, rather than $2s$ states, of N and B atoms,
respectively. The former (the occupied states) corresponds to the
$\sigma$ bonds between the constitute atoms in virtue of the
$sp^{2}$ hybridization, which guarantees the basic honeycomb
framework, high Young's modulus and tensile
strength.\cite{Chopra2} In contrast, the latter (the unoccupied
states) mainly determines some physical and chemical properties of
BNNTs (such as conductivity and chemical activity). The further
composition analysis on the corresponding frontier molecular
orbitals (FMOs) demonstrates that these occupied and unoccupied
states near the Fermi level exhibit quite different localized
characteristics. For instance, as shown in Fig.~\ref{fig:fig1}(b),
the highest occupied molecular orbital (HOMO) corresponds to
isolated electron pairs localized at the N atoms and has the
spindle-shape scheme like as $p_{z}$ orbitals, whereas the lowest
unoccupied molecular orbital (LUMO) is present as $\pi$ states
localized at the B-N pair along the tube-axis and is contributed
by the B-$2p$ (major) and N-$2p$ (minor) states. The corresponding
spatial orientations are well perpendicular to the cylindrical
surface. In principle, the nature of localized states near the
Fermi level is crucial for understanding the structures and
properties of doped BNNTs. In the following text, we
systematically study the adsorption behavior of single first-row
atoms on BNNTs and thoroughly probe the related adsorption
mechanism.

In the doping procedure, the interactions between host BNNTs and
foreign species can be usually classified into the physical and
chemical interactions. Indeed the chemical interaction might play
more significant role than the physical interaction in practical
applications of BNNTs, such as hydrogen storage\cite{Bando} and
modification of the electronic property.\cite{Tangcc} Since the
occupied and unoccupied states near the Fermi level in BNNTs have
the distinctly different localized characteristics, we could
deduce that the chemical interaction between BNNTs and dopants
certainly will be characterized as highly selective, and the
resulting adsorption behavior and effect ought to be more
complicated and interesting than those in CNTs in the same doping
procedure.

In general, the chemical activity of single atom could be
evaluated by the PE or FIP.\cite{Pauling} To more comprehensively
and systematically probe the mechanism of selective adsorption of
BNNTs, we choose three types of first-row atoms, Li, C and F as
the representatives. The order of PE is:
Li(0.98)$<$B(2.04)$<$C(2.55)$<$N(3.04)$<$F(3.98), and that of FIP
is: Li(5.392)$<$B(8.298)$<$C(11.260)$<$N(14.534)$<$F (17.422),
respectively. From the spatial orientations and distributions of
FMOs in BNNTs, we choose the top sites of B and N atoms, the
bridge sites between B and N atoms and the center site above the
hexagon as the initial adsorption sites. After full relaxations,
we obtain the optimized adsorption site, adsorption
energy\cite{ads} and adsorption height as summarized in
Table~\ref{tab:table1}.

\begin{table}
\caption{\label{tab:table1}Optimized adsorption sites, and
adsorption energies $E_{ads}$ and adsorption heights $H_{ads}$ of
first-row atoms (Li, C and F) on the single-walled (8,0) BNNT.}
\begin{ruledtabular}
\begin{tabular}{cccc}
 Atom& Adsorption Site& $E_{ads}$ (eV)& $H_{ads}$ (\AA)\\
\hline
\\[0.2ex]
&B&$0.44$&3.02\\[0.5ex]
Li&N&$0.32$&2.28 \\[0.5ex]
&Center\footnote{Above the hexagon}&$0.26$&1.80 \\[2ex]
{C}&Bridge\footnote{Perpendicular to the tube-axis}&$-2.00$&1.26 \\[2ex]
{F}&B&$-2.51$&1.84\\[0.2ex]
\end{tabular}
\end{ruledtabular}
\end{table}

\begin{figure*} [tbp]
\includegraphics[width=1.0\textwidth]{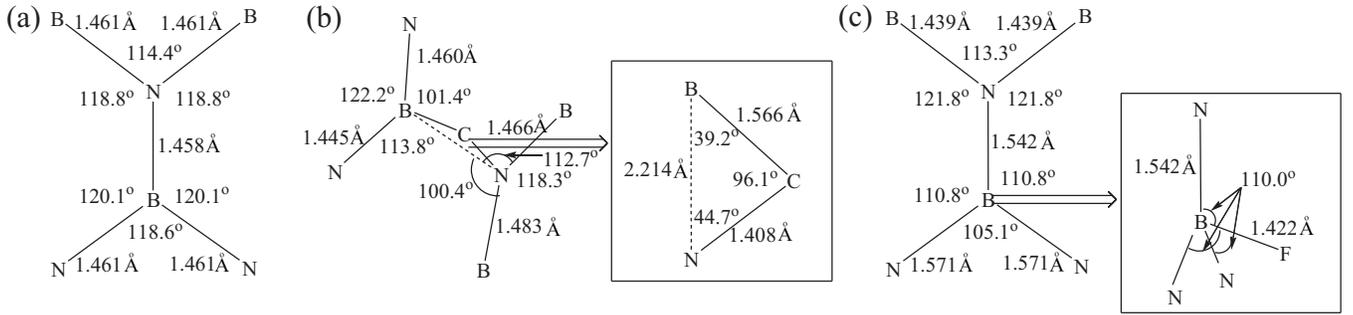}
\caption{\label{fig:fig2}Optimized morphologies of (a) the pure
(8,0) BNNT, (b) C/BNNT and (c) F/BNNT in the adsorption regions.
The bond lengths and angles between adsorbed atoms and host tubes
are shown in the corresponding insets.}
\end{figure*}

It is found that the F atom prefers to stay at the top site of the
B atom (see Fig.~\ref{fig:fig2}), and it is energetically
favorable for the C atom to stay at the B-N bridge site which is
perpendicular to the tube-axis (see Fig.~\ref{fig:fig2}). Whereas
the Li atom can hardly be adsorbed on the BNNT since its
adsorption energy is positive. The positive adsorption energy
indicates an endothermal reaction, which is in accordance with the
previous theoretical study.\cite{ZhouZh} The reason is that the
BNNT with partial ionic bonding is an electron-sufficient system,
and hardly accepts the excessive electrons from the environment.
Furthermore, the adsorption energy of the F atom on BNNT (F/BNNT)
is larger than that of the C atom on BNNT (C/BNNT) because the
electronegativity of the F atom is larger than that of the C atom.
The adsorption of the F and C atoms corresponds to a classical
chemical adsorption (or interaction). The formation of new
chemical bonds definitely changes the original electronic states,
and then influences the electronic properties of BNNTs.

Firstly, the chemical adsorption will induce a slight deformation
of the tube. The optimized bond lengths and angles in the
adsorption region (ADR) of C/BNNT and F/BNNT are displayed in
Fig.~\ref{fig:fig2}. In the C/BNNT, the C atom pushes the two
bonding B and N atoms along the bridge apart. The length of the
B-N bond adhered to the C atom (shown as the dashed line in
Fig.~\ref{fig:fig2}(b)) is changed from 1.461 {\AA} to 2.214
{\AA}, which indicates this B-N bond is weakened, and the
variations of other bonds are in the range of 0.04 {\AA}.
Furthermore, the N-B-N angles are changed from 120.1$^{\circ}$,
120.1$^{\circ}$ and 118.6$^{\circ}$ to 122.2$^{\circ}$,
101.4$^{\circ}$ and 113.8$^{\circ}$, respectively. And the B-N-B
angles are changed from 118.8$^{\circ}$, 118.8$^{\circ}$ and
114.4$^{\circ}$ to 118.3$^{\circ}$, 100.4$^{\circ}$ and
112.7$^{\circ}$, respectively. As shown in the inset of
Fig.~\ref{fig:fig2}(b), the bond angles of C-N-B and C-B-N are
44.7$^{\circ}$ and 39.2$^{\circ}$, and the lengths of C-N and C-B
bonds are 1.408 and 1.566 {\AA}, respectively. In the F/BNNT,
since the F atom is only adsorbed on the top of the B atom, the
induced effect on the tube's morphology mainly occurs in the
vicinity of the B atom. In detail, the N-B-N angles decrease to
110.8$^{\circ}$, 110.8$^{\circ}$ and 105.1$^{\circ}$,
respectively. And the B-N-B angles are changed to 121.8$^{\circ}$,
121.8$^{\circ}$ and 113.3$^{\circ}$, respectively.
Correspondingly, the first neighbor B-N bond lengths are changed
from 1.458 {\AA} and 1.461 {\AA} to 1.542 {\AA} and 1.571 {\AA},
and others are slightly changed with the magnitude of 0.03 {\AA}.
Especially, the B-F bond length is 1.422 {\AA}, which is somewhat
larger than the expected value for the B-F single bond (1.37
{\AA})\cite{Pauling} and the bond lengths of boron trifluoride
molecule (1.31 {\AA}).\cite{BF3} All F-B-N bond angles are
110.0$^{\circ}$ as shown in the inset of Fig.~\ref{fig:fig2}(c).
These changes of bond length and bond angle indicate that the
electronic states of the B and N atoms adjacent to the adsorbate
exhibit the mixed characteristic of the $sp^2$ and $sp^3$
hybridizations, instead of the $sp^2$ hybridization in the pure
BNNT, and the $sp^3$ component in the ADR of the F/BNNT is much
more than that of the C/BNNT.

\begin{figure} [tbp]
\includegraphics[width=0.5\textwidth]{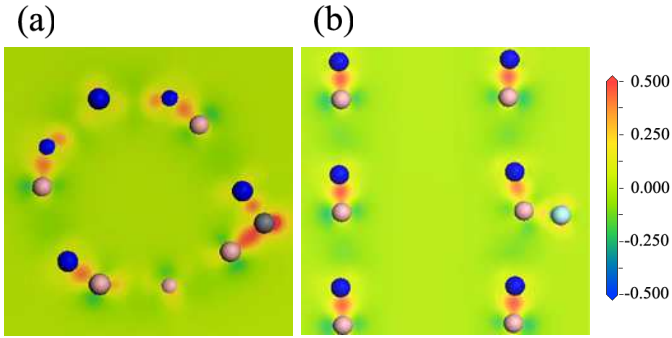}
\caption{\label{fig:fig3}(Color online) Deformation charge
densities (in units of a.u.) of a plane through the adsorbate and
the first neighbor B and N atoms of (a) the C/BNNT and (b) F/BNNT.
Blue, pink, grey and cyan balls correspond to N, B, C and F atoms,
respectively.}
\end{figure}

The more detailed analysis of the adsorption effect could be
obtained from the change of electronic states. Fig.~\ref{fig:fig3}
shows deformation charge densities of the C/BNNT and F/BNNT. It
can be seen that the B-N bond adjacent to the adsorbed C or F atom
is evidently weakened with the formation of new chemical bonds
between the C (or F) and B (and N) atoms. The Mulliken population
analysis\cite{Mulliken} further verifies that the adsorption not
only induces the charge transfer between the B, N and C atoms in
the ADR, but also leads to the charge re-distribution of two other
neighboring B (or N) atoms in the C/BNNT. Typically, the C atom
accepts the electrons from the nearest neighbor B atom, and
simultaneously donates the electrons to the nearest neighbor N
atom. And the net charge transfer between the C atom and BNNT does
not exceed 0.040 e. This reveals the C-N and C-B bonds are
covalent-like. However, in the F/BNNT, the adsorption occurs only
at the top of the B atom, so the charge transfer induced mainly
occurs between the B and F atoms, and the N atom is less involved
in. In detail, one F atom accepts 0.353 e from the B atom, which
is three times larger than the charge transfer between the B and N
atoms in the pure BNNT, and the B-F bond exhibits ionicity.
Simultaneously, the F adsorption further leads to slight charge
changes for two other neighboring N atoms and their neighboring B
atoms (0.020 e/N atom and 0.035 e/B atom, respectively).

\begin{figure} [tbp]
\includegraphics[width=0.5\textwidth]{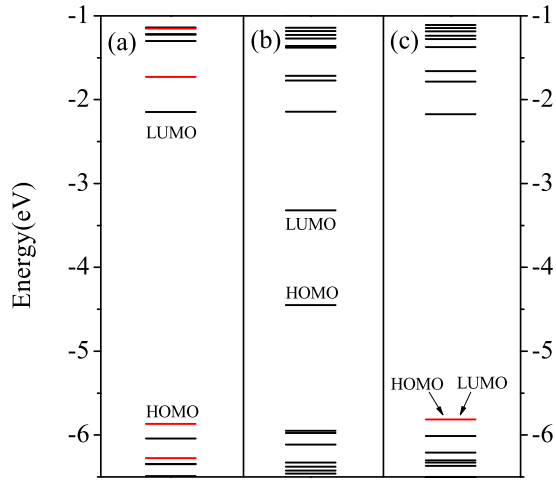}
\caption{\label{fig:fig4}(Color online) Energy levels of (a) the
pure (8,0) BNNT, (b) C/BNNT and (c) F/BNNT. The HOMO and LUMO are
denoted. Red line means the degenerated level.}
\end{figure}

\begin{figure} [tbp]
\includegraphics[width=0.5\textwidth]{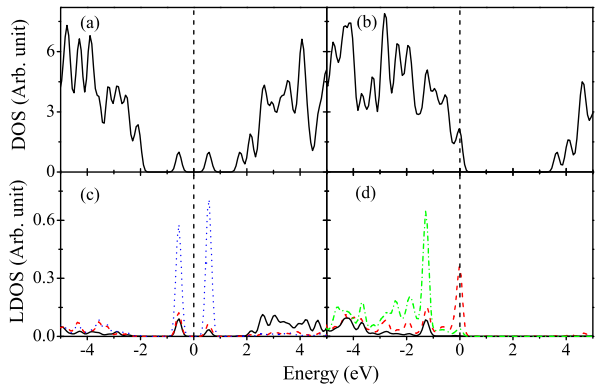}
\caption{\label{fig:fig5}(Color online) (a) Density of states (DOS)
and (c) LDOS for the C/BNNT; and (b) DOS and (d) LDOS for the
F/BNNT. Black solid, red dashed, blue dotted and green dash-dotted
lines correspond to B, N, C and F atoms, respectively. The Fermi
level is set to zero energy.}
\end{figure}

The above results can be better elucidated based on the FMO
theory. Due to its strongest oxidizability, the F atom in the
F/BNNT tends to directly interact with the B atom (low
electronegativity) rather than the N atom (high
electronegativity). The formation of chemical bond are originated
from the interaction between the LUMO of the F atom and the FMOs
of about 2.10 eV below the Fermi level (which are partially
contributed by B-$2p$ states as shown in Fig.~\ref{fig:fig1}(a))
of the BNNT. It is interesting to note that the HOMO of the BNNT
does not really contribute to the chemical bond mentioned above.
And the related detail could be illustrated by the LDOS of the
F/BNNT in Fig.~\ref{fig:fig5}(d). On the other hand, the
electronegativity of the C atom is between these of the B and N
atoms, and thus the HOMO (LUMO) of the C atom interacts with the
LUMO (HOMO) of the BNNT when the C atom is close to the tube. As
illustrated in Fig.~\ref{fig:fig1}(b), the LUMO and HOMO of the
BNNT are mainly contributed by B-$2p$ and N-$2p$ states,
respectively. Therefore, the C atom is adsorbed at the bridge site
which is perpendicular to the tube-axis. The resulting energy
levels of the C/BNNT and F/BNNT are shown in Fig.~\ref{fig:fig4}.
The gap between the HOMO and LUMO of the C/BNNT ($1.13$ eV) is
smaller than that of the pure BNNT ($3.72$ eV) due to the
introduction of donor and acceptor states (as shown in
Fig.~\ref{fig:fig5}(a)). The calculated LDOS
(Fig.~\ref{fig:fig5}(c)) shows that the HOMO is originated from
the orbitals of C, B and N atoms, but the LUMO is mainly
contributed by the orbitals of the C atom. This feature indicates
that if we increase the content of C atoms in C-BNNTs, the gap
will decrease and a transition from semiconducting to metallic
behavior might occur in some content of C atoms. This is in
accordance with the previous study on B-C-N nanotubes.\cite{Blase}
In the F/BNNT, the HOMO and LUMO are degenerate (as shown in
Fig.~\ref{fig:fig5}(b)), which suggests a conducting behavior.
Fig.~\ref{fig:fig5}(d) shows that the HOMO and LUMO are
contributed by the orbitals of N atoms (major contribution) and
the F atom (minor contribution). The LDOS of the F atom is only
significant much below the Fermi level. This further verifies the
chemical interaction between the F atom and tube is typically
ionic. Since the charge does unambiguously transfer from the BNNT
to the more electronegative F atom, the Fermi level is evidently
shifted down into the original completely occupied states
contributed by the N atoms. This means, in this way, the F doping
system will exhibit metallic behavior, which is in agreement with
the experiment of fluorination of BNNTs by Tang {\it et
al.}.\cite{Tangcc}

Similar to the case of BNNTs, the localized states at the valence
and conduction band edges of other compound nanotubes, such as SiC
nanotubes, are also distinctly different. They would play an
important role on the physical and chemical properties (such as
conductivity and adsorption property) of tubes. Modulating the
{\it localized} and {\it delocalized} characteristics of
electronic states by adsorption or doping could improve the
potential of tubes as functional substrates for some applications
to the utmost extent.

\section{Conclusions}
Using first-principles calculations, we study the adsorption of
first-row atoms with different electronegativity (Li, C and F) on
zigzag single-walled BNNTs. We observe that adsorption exhibits
interesting selectivity. Typically, the F atom prefers to adsorb
on the top of the B atom, the C atom is energetically favorable to
stay on the bridge site which is perpendicular to the tube-axis,
and the Li atom hardly adheres to the tube, corresponding to an
endothermal reaction. This is related to the localized
characteristics of occupied and unoccupied states near the Fermi
level of host BNNTs and the electronegativity of adsorbate with
respect to the electronegativity of B and N atoms in tubes.
Furthermore, the adsorption of C and F atoms on BNNTs would lead
to different changes on the electronic properties. The former
results in the decrease of energy gap of the BNNT, and the latter
leads to a transition from semiconducting to metallic behavior. It
is expected that the electronic properties of BNNTs could be
tailored via certain adsorption.

This work is supported by the National Natural Science Foundation
of China (Grant Nos. 10325415 and 10404016), and Ministry of
Education of China (Grant No. SRFDP 20050003085).


\begin{references}
\bibitem{ads} The adsorption energies $E_{ads}$ are calculated
from the expression
$E_{ads}=E_{b}[\mbox{A/BNNT}]-E_{b}[\mbox{BNNT}]$, where
$E_{b}[\mbox{BNNT}]$ and $E_{b}[\mbox{A/BNNT}]$ are the binding
energies of the optimized pure BNNT and A/BNNT, respectively.
\end{references}
\end{document}